\documentclass{aa}  

\usepackage{graphicx}
\usepackage{txfonts}
\usepackage{color}
\usepackage{stfloats}
\usepackage{multirow}
\usepackage{mhchem}
\usepackage{booktabs}

\newcommand{\tablenotea}[1]{\parbox{14.8cm}{\indent \footnotesize{#1}}}
\newcommand{\tablenoteb}[1]{\parbox{8.8cm}{\indent \footnotesize{#1}}}

\newcommand{\cpl}{Chem. Phys. Lett.}
\newcommand{\chemrev}{Chem. Rev.}

\newcommand{\jcc}{J. Comp. Chem.}

\newcommand{\jms}{J. Mol. Spectr.}

\newcommand{\jpca}{J. Phys. Chem. A}

\newcommand{\natastro}{Nat. Astron.}
\newcommand{\pccp}{Phys. Chem. Chem. Phys.}

\begin{document}

\title{Detection of thioacetaldehyde (CH$_3$CHS) in \mbox{TMC-1}:\\ sulfur-oxygen differentiation along the hydrogenation sequence\thanks{Based on observations carried out with the Yebes 40m telescope (projects 19A003, 20A014, 20D023, 21A011, 21D005, and 23A024). The 40m radio telescope at Yebes Observatory is operated by the Spanish Geographic Institute (IGN; Ministerio de Transportes y Movilidad Sostenible).}}

\titlerunning{CH$_3$CHS in \mbox{TMC-1}}
\authorrunning{Ag\'undez et al.}

\author{M.~Ag\'undez\inst{1}, G.~Molpeceres\inst{1}, C.~Cabezas\inst{1}, N.~Marcelino\inst{2,3}, B.~Tercero\inst{2,3}, R.~Fuentetaja\inst{1}, P.~de~Vicente\inst{3}, \and J.~Cernicharo\inst{1}}

\institute{
Instituto de F\'isica Fundamental, CSIC, Calle Serrano 123, E-28006 Madrid, Spain\\ \email{marcelino.agundez@csic.es, jose.cernicharo@csic.es} \and
Observatorio Astron\'omico Nacional, IGN, Calle Alfonso XII 3, E-28014 Madrid, Spain \and
Observatorio de Yebes, IGN, Cerro de la Palera s/n, E-19141 Yebes, Guadalajara, Spain
}

\date{Received; accepted}

% \abstract{}{}{}{}{} 
% 5 {} token are mandatory
 
\abstract
% context heading (optional), leave it empty if necessary
{In recent years the chemistry of sulfur in the interstellar medium has experienced a renewed interest due to the detection of a large variety of molecules containing sulfur. Here we report the first identification in space of a new S-bearing molecule, thioacetaldehyde (CH$_3$CHS), which is the sulfur counterpart of acetaldehyde (CH$_3$CHO). The astronomical observations are part of QUIJOTE, a Yebes\,40m Q band line survey of the cold dense cloud \mbox{TMC-1}. We detected seven individual lines corresponding to $A$ and $E$ components of the four most favorable rotational transitions of CH$_3$CHS covered in the Q band (31.0-50.3 GHz). Assuming a rotational temperature of 9 K, we derive a column density of 9.8\,$\times$\,10$^{10}$ cm$^{-2}$ for CH$_3$CHS, which implies that it is 36 times less abundant than its oxygen counterpart CH$_3$CHO. By comparing the column densities of the O- and S-bearing molecules detected in \mbox{TMC-1}, we find that as molecules increase their degree of hydrogenation, sulfur-bearing molecules become less abundant compared to their oxygen analog. That is, hydrogenation seems to be less favored for S-bearing molecules than for O-bearing ones in cold sources like \mbox{TMC-1}. We explored potential formation pathways to CH$_3$CHS and implemented them into a chemical model, which however underestimates by several orders of magnitude the observed abundance of thioacetaldehyde. Quantum chemical calculations carried out for one of the potential formation pathways, the S + C$_2$H$_5$ reaction, indicate that formation of CH$_3$CHS is only a minor channel in this reaction.}


\maketitle

\section{Introduction}

There are still many gaps of knowledge on the chemistry of sulfur in interstellar space. This is well illustrated by the missing sulfur problem, which states that most of the sulfur budget in cold dense clouds is still pending characterization \citep{Ruffle1999,Martin-Domenech2016,Vidal2017,Navarro-Almaida2020}. In the last years, many new sulfur-bearing molecules have been reported in dense clouds, namely, S$_2$H \citep{Fuente2017}, HCS, HSC \citep{Agundez2018}, NS$^+$, HC$_3$S$^+$, NCS, HCCS, H$_2$CCS, H$_2$CCCS, C$_4$S, C$_5$S, HCSCN, HCSCCH \citep{Cernicharo2018,Cernicharo2021a,Cernicharo2021b,Cernicharo2021c}, HCOSH \citep{Rodriguez-Almeida2021}, HCCS$^+$ \citep{Cabezas2022}, HC$_4$S \citep{Fuentetaja2022}, HSO \citep{Marcelino2023}, HNSO \citep{Sanz-Novo2024}, NCCHCS \citep{Cabezas2024}, HCNS, NC$_3$S, HC$_3$S \citep{Cernicharo2024a,Cernicharo2024b}, and surprisingly the metal-bearing molecules MgS, and NaS \citep{Rey-Montejo2024}. However, none of the above molecules are a major reservoir of sulfur. For example, in \mbox{TMC-1}, if we adopt a column density of H$_2$ of 10$^{22}$ cm$^{-2}$ \citep{Cernicharo1987}, the inventory of S-bearing molecules detected according to our records (see below in Sect.\,\ref{sec:discussion}) accounts for only 0.15\,\% of the cosmic abundance of sulfur \citep{Asplund2009}, meaning that 99.85\,\% is still missing, awaiting identification.

The large number of S-bearing molecules detected did not help to solve the missing sulfur problem, but it has allowed to put exquisite observational constraints on the chemistry of this element, in particular in cold dense clouds. Although many of the sulfurated molecules detected have their oxygenated counterpart detected as well, there are significant differences between the chemistries of sulfur and oxygen. For example, the carbon chains C$_2$S, C$_3$S, and C$_5$S are far more abundant than their oxygen counterparts \citep{Cernicharo2021b,Cernicharo2021d}. The hydrogenated O-bearing molecules HCO, H$_2$CO, HCCO, CH$_2$CO, HC$_3$O, and HCCCHO have their sulfur counterpart detected \citep{Agundez2013,Loison2016,Cernicharo2020a,Cernicharo2021b,Cernicharo2021c,Cernicharo2021d,Cernicharo2024b}, although in some cases the oxygen molecule is more abundant than the sulfur one, while in other cases the contrary is found. There are also many O-bearing complex organic molecules for which the sulfur counterpart remains elusive. These are C$_2$H$_3$OH, C$_2$H$_5$OH, CH$_3$OCH$_3$, HCOOH, HCOOCH$_3$, HC$_5$O, HC$_7$O, $c$-H$_2$C$_3$O, C$_2$H$_3$CHO, CH$_3$COCH$_3$, and C$_2$H$_5$CHO \citep{Loison2016,McGuire2017,Cordiner2017,Cernicharo2020a,Agundez2021,Agundez2023a}.

In this Letter we report the detection in the cold dense cloud \mbox{TMC-1} of thioacetaldehyde (CH$_3$CHS), the sulfur analog of acetaldehyde (CH$_3$CHO). Acetaldehyde is a well known interstellar molecule that was detected in space in the early years of radioastronomy \citep{Gottlieb1973}. It is known to be present in \mbox{TMC-1} \citep{Matthews1985} and its chemistry is relatively well understood \citep{Lamberts2019,Vazart2020,Enrique-Romero2021,Fedoseev2022,Ferrero2022,Molpeceres2022a}. However, its sulfur cousin CH$_3$CHS has so far eluded detection in interstellar space \citep{Margules2020} and very little is known on its chemistry.

\section{Astronomical observations}

The astronomical observations on which this work is based belong to the ongoing Yebes\,40m Q-band line survey of \mbox{TMC-1,} QUIJOTE (Q-band Ultrasensitive Inspection Journey to the Obscure TMC-1 Environment; \citealt{Cernicharo2021e}). Briefly, QUIJOTE is a Q-band line survey (31.0-50.3 GHz) of the cold dense cloud \mbox{TMC-1} at the position of the cyanopolyyne peak ($\alpha_{J2000}=4^{\rm h} 41^{\rm  m} 41.9^{\rm s}$ and $\delta_{J2000}=+25^\circ 41' 27.0''$). The observations were carried out in the frequency-switching observing mode, with a frequency throw of either 8 or 10 MHz. We used a 7 mm dual linear polarization receiver connected to a set of eight fast Fourier transform spectrometer, which allows to cover the full Q band in one shot with a spectral resolution of 38.15 kHz in both polarizations \citep{Tercero2021}. The intensity scale is the antenna temperature, $T_A^*$, which has an estimated uncertainty due to calibration of 10\,\%. The main beam brightness temperature, $T_{\rm mb}$, can be obtained by dividing $T_A^*$ by $B_{\rm eff}$/$F_{\rm eff}$ (see note in Table\,\ref{table:lines}). The half power beam width (HPBW) can be approximated as HPBW($''$) = 1763/$\nu$(GHz). We used the latest QUIJOTE dataset, which includes observations carried out between November 2019 and July 2024. The total on-source telescope time is 1509.2 h, of which 736.6 h correspond to a frequency throw of 8 MHz and 772.6 h to a throw of 10 MHz. The $T_A^*$ rms noise level varies between 0.06 mK at 32 GHz and 0.18 mK at 49.5 GHz. The procedure used to reduce and analyze the data is described in \cite{Cernicharo2022}. All data were analyzed using the GILDAS software\footnote{\texttt{https://www.iram.fr/IRAMFR/GILDAS/}}.

\begin{table*}
\small
\caption{Observed line parameters of CH$_3$CHS in \mbox{TMC-1}.}
\label{table:lines}
\centering
\begin{tabular}{lrlcccccc}
\hline \hline
 \multicolumn{1}{l}{Transition} & \multicolumn{1}{c}{$E_{\rm up}$} & \multicolumn{1}{c}{$\nu_{\rm calc}$} & \multicolumn{1}{c}{$T_A^*$ peak} & \multicolumn{1}{c}{$\Delta v$\,$^a$}      & \multicolumn{1}{c}{$V_{\rm LSR}$}      & \multicolumn{1}{c}{$\int T_A^* dv$} & \multicolumn{1}{c}{S/N\,$^b$} & \multicolumn{1}{c}{$B_{\rm eff}$/$F_{\rm eff}$\,$^c$} \\
& \multicolumn{1}{c}{(K)}        & \multicolumn{1}{c}{(MHz)}        & \multicolumn{1}{c}{(mK)}                   & \multicolumn{1}{c}{(km s$^{-1}$)}  & \multicolumn{1}{c}{(km s$^{-1}$)}  & \multicolumn{1}{c}{(mK km s$^{-1}$)} & \multicolumn{1}{c}{($\sigma$)} & \\
\hline
3$_{0,3}$-2$_{0,2}$ $E$ & 3.2 & 33160.973\,$^d$ & -- & -- & -- & -- & -- & 0.66 \\ % blended with stronger line, fit is not attempted
3$_{0,3}$-2$_{0,2}$ $A$ & 3.2 & 33161.660 & $0.44 \pm 0.10$ & $1.02 \pm 0.20$ & $5.70 \pm 0.30$ & $0.48 \pm 0.16$ &  8.1 & 0.66 \\
4$_{1,4}$-3$_{1,3}$ $A$ & 7.3 & 43340.053 & $0.53 \pm 0.13$ & $0.83 \pm 0.28$ & $6.24 \pm 0.28$ & $0.46 \pm 0.18$ &  7.6 & 0.57 \\
4$_{1,4}$-3$_{1,3}$ $E$ & 7.3 & 43340.923 & $0.46 \pm 0.13$ & $0.83 \pm 0.30$ & $5.82 \pm 0.30$ & $0.41 \pm 0.20$ &  6.7 & 0.57 \\
4$_{0,4}$-3$_{0,3}$ $E$ & 5.3 & 44198.358 & $0.31 \pm 0.11$ & $1.16 \pm 0.38$ & $6.00 \pm 0.26$ & $0.38 \pm 0.16$ &  6.3 & 0.56 \\
4$_{0,4}$-3$_{0,3}$ $A$ & 5.3 & 44199.272 & $0.49 \pm 0.11$ & $0.31 \pm 0.15$ & $5.63 \pm 0.18$ & $0.16 \pm 0.10$ &  5.1 & 0.56 \\
4$_{1,3}$-3$_{1,2}$ $E$ & 7.5 & 45111.102 & $0.48 \pm 0.17$ & $0.49 \pm 0.18$ & $5.86 \pm 0.13$ & $0.25 \pm 0.11$ &  4.2 & 0.55 \\
4$_{1,3}$-3$_{1,2}$ $A$ & 7.5 & 45113.545 & $0.33 \pm 0.17$ & $0.71 \pm 0.35$ & $5.99 \pm 0.25$ & $0.25 \pm 0.17$ &  3.5 & 0.55 \\
\hline
\end{tabular}
\tablenotea{\\
\textbf{Note.} The line parameters $T_A^*$ peak, $\Delta v$, $V_{\rm LSR}$, and $\int T_A^* dv$ and the associated errors are derived from a Gaussian fit to each line profile. $^a$\,$\Delta v$ is the full width at half maximum (FWHM). $^b$\,The signal-to-noise ratio is computed as S/N = $\int T_A^* dv$ / [rms $\times$ $\sqrt{\Delta v \times \delta \nu (c / \nu_{\rm calc})}$], where $c$ is the speed of light, $\delta \nu$ is the spectral resolution (0.03815 MHz), the rms is given in the uncertainty of $T_A^*$ peak, and the rest of parameters are given in the table. $^c$\,$B_{\rm eff}$ is given by the Ruze formula $B_{\rm eff}$\,=\,0.797\,$\exp{[-(\nu/71.1)^2]}$, where $\nu$ is the frequency in GHz and $F_{\rm eff}$\,=\,0.97. $^d$\,Line is blended with the stronger CH$_3$C$^{13}$CH line 2$_0$-1$_0$ at 33160.940 MHz. No line fit is attempted.
}
\end{table*}

\section{Results} \label{sec:results}

Thioacetaldehyde (CH$_3$CHS) is an asymmetric rotor in which the internal rotation of the CH$_3$ group, which has C$_{3v}$ symmetry, couples to the overall rotation of the molecule. As a result, the rotational levels split into $A$ and $E$ substates and the molecule can exist in different torsional states. All lines observed here belong to the ground torsional state $v_t$\,=\,0. The next torsional state $v_t$\,=\,1 lies 229 K above the ground torsional state and is therefore not populated at the cold temperatures of \mbox{TMC-1}. The rotational spectrum of thioacetaldehyde was measured in the laboratory in the microwave range (7-41 GHz) by \cite{Kroto1974} and \cite{Kroto1976}, and more recently at millimeter and submillimeter waves (150-660 GHz) by \cite{Margules2020}. The total dipole moment was measured to be 2.33\,$\pm$\,0.02 D, with components along the $a$ and $b$ axes of 2.26\,$\pm$\,0.02 D and 0.56\,$\pm$\,0.01 D, respectively \citep{Kroto1976}. All transitions observed here are $a$-type. We adopted the rotational spectroscopy of CH$_3$CHS from the Lille Spectroscopic Database\footnote{\texttt{https://lsd.univ-lille.fr/}}, which is based in the aforementioned references.

\begin{figure}
\centering
\includegraphics[angle=0,width=\columnwidth]{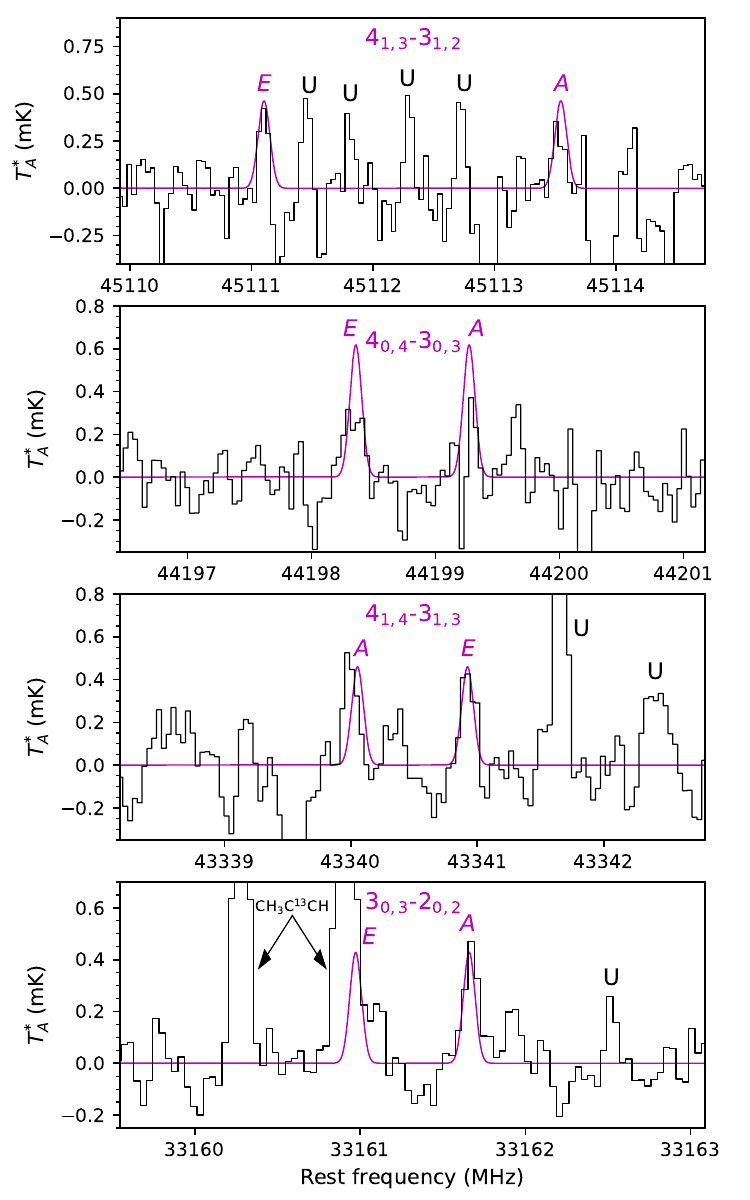}
\caption{Lines of CH$_3$CHS observed in \mbox{TMC-1} (black histogram). Negative features correspond to artifacts caused by the frequency-switching technique. The line parameters are given in Table~\ref{table:lines}. In color magenta we show the LTE calculated line profiles adopting a column density of 9.8\,$\times$\,10$^{10}$\,cm$^{-2}$, a rotational temperature of 9.0\,K, a full width at half maximum of 0.76 km s$^{-1}$ (the average of the values observed for the seven detected components; see Table\,\ref{table:lines}), and a circular emission distribution with a diameter of 80\,$''$.} \label{fig:lines}
\end{figure}

For a rotational temperature of 9 K, which is the gas kinetic temperature in \mbox{TMC-1} \citep{Agundez2023b}, the most intense predicted lines of CH$_3$CHS in the Q band are the 3$_{0,3}$-2$_{0,2}$ at 33.16 GHz, the 4$_{1,4}$-3$_{1,3}$ at 43.34 GHz, the 4$_{0,4}$-3$_{0,3}$ at 44.20 GHz, and the 4$_{1,3}$-3$_{1,2}$ at 45.11 GHz, each of them consisting of a doublet of two components $A$ and $E$ separated by 0.6-2.4 MHz and with nearly identical intensities. We detect these four doublets of lines in the Q band scan of \mbox{TMC-1}. The line parameters obtained from a Gaussian fit to the line profiles are given in Table\,\ref{table:lines} and the lines are shown in Fig.\,\ref{fig:lines}. It is seen that the lines are weak, with intensities well below 1 mK in antenna temperature, although they are detected with signal-to-noise ratios above 3\,$\sigma$, in the range 3.5-8.1\,$\sigma$ (see Table\,\ref{table:lines}). The $E$ component of the 3$_{0,3}$-2$_{0,2}$ transition at 33160.973 MHz lies very close in frequency to a brighter line, the 2$_0$-1$_0$ transition of CH$_3$C$^{13}$CH at 33160.940 MHz (Cernicharo et al., in preparation), and thus cannot be adequately fitted. The seven fitted components are centered at the systemic velocity of \mbox{TMC-1}, 5.83 km s$^{-1}$ \citep{Cernicharo2020b}. The $A$/$E$ doublets of the transitions 3$_{1,3}$-2$_{1,2}$ at 32.51 GHz and 3$_{1,2}$-2$_{1,1}$ at 33.83 GHz are predicted with about half of the intensity of the four doublets of lines shown in Fig.\,\ref{fig:lines}. They are not clearly detected above the noise level, which is consistent with their low predicted intensities. Although the signal-to-noise ratios of the lines detected are moderately low, the availability of the four brightest doublets in the Q band and the fact that no bright line is missing makes us to be secure about the detection of thioacetaldehyde in \mbox{TMC-1}.

In order to determine the column density of thioacetaldehyde in \mbox{TMC-1} we have assumed Local Thermodynamic Equilibrium (LTE). The narrow range of upper level energies of the seven components detected (3.2-7.5 K) and the sizable errors in the velocity-integrated intensities do not allow to constrain precisely the rotational temperature. We thus adopted a rotational temperature of 9 K, the gas kinetic temperature in \mbox{TMC-1} \citep{Agundez2023b}, which is consistent with the observed relative intensities, and an emission distribution consisting of a circle with a diameter of 80\,$''$, which is consistent with the emission size of most molecules mapped in \mbox{TMC-1} \citep{Cernicharo2023}. We derive a column density of 9.8\,$\times$\,10$^{10}$ cm$^{-2}$ for CH$_3$CHS in \mbox{TMC-1}. The line profiles calculated under LTE for that column density are compared with the observed ones in Fig.\,\ref{fig:lines}. A variation of 1 K in the adopted rotational temperature has only a moderate impact of just 5\,\% on the value of the column density determined.

\section{Discussion} \label{sec:discussion}

\begin{table}[ht!]
\small
\caption{Observed column densities of O- and S-bearing molecules in \mbox{TMC-1}.}
\label{table:column_densities}
\centering
\begin{tabular}{lrrlrr}
\hline \hline
\multicolumn{1}{l}{Molecule} & \multicolumn{1}{c}{$N$ (cm$^{-2}$)} & \multicolumn{1}{c}{Ref.} & \multicolumn{1}{c}{Molecule} & \multicolumn{1}{c}{$N$ (cm$^{-2}$)} & \multicolumn{1}{c}{Ref.}\\
 \hline
%==================================================
  OH              & 3.0\,$\times$\,10$^{15}$ &  (1) &             &                          &      \\
                  &                          &      & H$_2$S      & 1.2\,$\times$\,10$^{14}$ &  (9) \\
                  &                          &      & SO          & 3.4\,$\times$\,10$^{13}$ & (10) \\
                  &                          &      & HSO         & 7.0\,$\times$\,10$^{10}$ & (11) \\
                  &                          &      & SO$_2$      & 3.0\,$\times$\,10$^{12}$ &  (1) \\
\hline
%==================================================
  CO              & 1.7\,$\times$\,10$^{18}$ &  (1) & CS          & 1.1\,$\times$\,10$^{14}$ & (12) \\
  HCO             & 1.1\,$\times$\,10$^{12}$ &  (2) & HCS         & 5.5\,$\times$\,10$^{12}$ & (13) \\
                  &                          &      & HSC         & 1.3\,$\times$\,10$^{11}$ & (13) \\
  HCO$^+$         & 9.3\,$\times$\,10$^{13}$ &  (1) & HCS$^+$     & 5.5\,$\times$\,10$^{12}$ & (12) \\
  H$_2$CO         & 5.0\,$\times$\,10$^{14}$ &  (1) & H$_2$CS     & 3.7\,$\times$\,10$^{13}$ & (12) \\
  CH$_3$OH        & 4.8\,$\times$\,10$^{13}$ &  (3) & CH$_3$SH    & 1.7\,$\times$\,10$^{12}$ & (14) \\
\hline
%==================================================
                  &                          &      & OCS         & 2.2\,$\times$\,10$^{13}$ &  (1) \\
  HCO$_2^+$       & 4.0\,$\times$\,10$^{11}$ &  (3) &             &                          &      \\
  HCOOH           & 1.4\,$\times$\,10$^{12}$ &  (3) &             &                          &      \\
\hline
%==================================================
  CCO             & 7.5\,$\times$\,10$^{11}$ &  (2) & CCS         & 3.4\,$\times$\,10$^{13}$ & (12) \\
                  &                          &      & HCCS$^+$    & 1.1\,$\times$\,10$^{12}$ & (15) \\
  HCCO            & 7.7\,$\times$\,10$^{11}$ &  (2) & HCCS        & 6.8\,$\times$\,10$^{11}$ & (13) \\
  CH$_2$CO        & 1.4\,$\times$\,10$^{13}$ &  (3) & CH$_2$CS    & 7.8\,$\times$\,10$^{11}$ & (13) \\
  CH$_3$CO$^+$    & 3.2\,$\times$\,10$^{11}$ &  (4) &             &                          &      \\
  CH$_3$CHO       & 3.5\,$\times$\,10$^{12}$ &  (3) & CH$_3$CHS   & 9.8\,$\times$\,10$^{10}$ & (14) \\
  C$_2$H$_3$OH    & 2.5\,$\times$\,10$^{12}$ &  (5) &             &                          &      \\
  C$_2$H$_5$OH    & 1.1\,$\times$\,10$^{12}$ &  (6) &             &                          &      \\
  CH$_3$OCH$_3$   & 2.5\,$\times$\,10$^{12}$ &  (5) &             &                          &      \\
  HCOOCH$_3$      & 1.1\,$\times$\,10$^{12}$ &  (5) &             &                          &      \\
  C$_3$O          & 1.2\,$\times$\,10$^{12}$ &  (2) & C$_3$S      & 6.8\,$\times$\,10$^{12}$ & (12) \\
  HC$_3$O         & 1.3\,$\times$\,10$^{11}$ &  (2) & HC$_3$S     & 1.5\,$\times$\,10$^{11}$ & (16) \\
  HC$_3$O$^+$     & 2.1\,$\times$\,10$^{11}$ &  (3) & HC$_3$S$^+$ & 4.0\,$\times$\,10$^{11}$ & (17) \\
  HCCCHO          & 1.5\,$\times$\,10$^{12}$ &  (7) & HCCCHS      & 3.2\,$\times$\,10$^{11}$ & (18) \\
  $c$-H$_2$C$_3$O & 4.0\,$\times$\,10$^{11}$ &  (3) &             &                          &      \\
                  &                          &      & H$_2$C$_3$S & 3.7\,$\times$\,10$^{11}$ & (13) \\
  C$_2$H$_3$CHO   & 2.2\,$\times$\,10$^{11}$ &  (5) &             &                          &      \\
  C$_2$H$_5$CHO   & 1.9\,$\times$\,10$^{11}$ &  (6) &             &                          &      \\
  CH$_3$COCH$_3$  & 1.4\,$\times$\,10$^{11}$ &  (6) &             &                          &      \\
                  &                          &      & C$_4$S      & 3.8\,$\times$\,10$^{10}$ & (12) \\
                  &                          &      & HC$_4$S     & 9.5\,$\times$\,10$^{10}$ & (19) \\
  C$_5$O          & 1.5\,$\times$\,10$^{10}$ &  (2) & C$_5$S      & 3.0\,$\times$\,10$^{10}$ & (12) \\
  HC$_5$O         & 1.4\,$\times$\,10$^{12}$ &  (2) &             &                          &      \\
  HC$_7$O         & 6.5\,$\times$\,10$^{11}$ &  (2) &             &                          &      \\
\hline
%==================================================
  NO              & 2.7\,$\times$\,10$^{14}$ &  (1) & NS          & 8.0\,$\times$\,10$^{12}$ &  (1) \\
                  &                          &      & NS$^+$      & 5.2\,$\times$\,10$^{10}$ & (20) \\
  NCO             & 7.4\,$\times$\,10$^{11}$ &  (8) & NCS         & 7.8\,$\times$\,10$^{11}$ & (13) \\
  HNCO            & 1.1\,$\times$\,10$^{13}$ &  (8) & HNCS        & 3.8\,$\times$\,10$^{11}$ & (13) \\
  HCNO            & 7.8\,$\times$\,10$^{10}$ &  (8) & HCNS        & 9.0\,$\times$\,10$^{9}$  & (21) \\
  HOCN            & 1.5\,$\times$\,10$^{11}$ &  (8) & HSCN        & 5.8\,$\times$\,10$^{11}$ & (13) \\
  H$_2$NCO$^+$    & 1.8\,$\times$\,10$^{10}$ &  (8) &             &                          &      \\
  HCOCN           & 3.5\,$\times$\,10$^{11}$ &  (7) & HCSCN       & 1.3\,$\times$\,10$^{12}$ & (22) \\
                  &                          &      & NC$_3$S     & 1.4\,$\times$\,10$^{11}$ & (15) \\
                  &                          &      & NCCHCS      & 1.2\,$\times$\,10$^{11}$ & (23) \\
%==================================================
\hline
\end{tabular}
\tablenoteb{\\
\textbf{References:}
(1) \cite{Agundez2013}; (2) \cite{Cernicharo2021d}; (3) \cite{Cernicharo2020a}; (4) \cite{Cernicharo2021f}; (5) \cite{Agundez2021}; (6) \cite{Agundez2023a}; (7) \cite{Cernicharo2021c}; (8) \cite{Cernicharo2024a}; (9) \cite{Navarro-Almaida2020}; (10) \cite{Loison2019}; (11) \cite{Marcelino2023}; (12) Fuentetaja et al. in preparation; (13) \cite{Cernicharo2021b}; (14) this work; (15) \cite{Cabezas2022}; (16) \cite{Cernicharo2024b}; (17) \cite{Cernicharo2021a}; (18) \cite{Cernicharo2021c}; (19) \cite{Fuentetaja2022}; (20) \cite{Cernicharo2018}; (21) \cite{Cernicharo2024a}; (22) \cite{Cernicharo2021c}; (23) \cite{Cabezas2024}.
}
\end{table}

The cold dense cloud \mbox{TMC-1} contains a remarkably rich variety of sulfur-containing molecules. Despite sulfur is $\sim$40 times less abundant than oxygen \citep{Asplund2009}, the number of S-bearing molecules identified to date in \mbox{TMC-1} is comparable to that of O-bearing ones, 33 and 39, respectively, according to our records (see Table\,\ref{table:column_densities}). It is worth to note that most of the S-bearing molecules detected in \mbox{TMC-1} have their oxygen counterpart detected as well, as illustrated in Table\,\ref{table:column_densities}. Those cases in which no oxygen counterpart is detected, such as HSC, H$_2$C$_3$S, C$_4$S, HC$_4$S, NS$^+$, NC$_3$S, and NCCHCS, reflect the existing chemical differences between sulfur and oxygen. For example, the metastable isomer HOC has not been observed in space, probably because its formation from H and CO is endothermic, unlike in the sulfur case \citep{Marenich2003,Puzzarini2005}. As discussed by \cite{Loison2016} and \cite{Shingledecker2019}, chemical kinetics explain the non-detection of the most stable C$_3$H$_2$O isomer, propadienone (H$_2$C$_3$O), and the detection of the two isomers higher in energy propynal (HCCCHO) and cyclopropenone ($c$-H$_2$C$_3$O), while in the sulfur case, the two most stable isomers (H$_2$C$_3$S and HCCCHS) are detected \citep{Cernicharo2021b,Cernicharo2021c}. Further differences are illustrated by the fact that the sulfur carbon chains C$_4$S, HC$_4$S, and NC$_3$S and the cation NS$^+$ have no oxygen counterpart detected \citep{Cernicharo2018,Cernicharo2021b,Cernicharo2021d,Cernicharo2024b,Fuentetaja2022}, while NCCHCS is more abundant than its oxygen analogue NCCHCO \citep{Cabezas2024}.

\begin{figure}
\centering
\includegraphics[angle=0,width=\columnwidth]{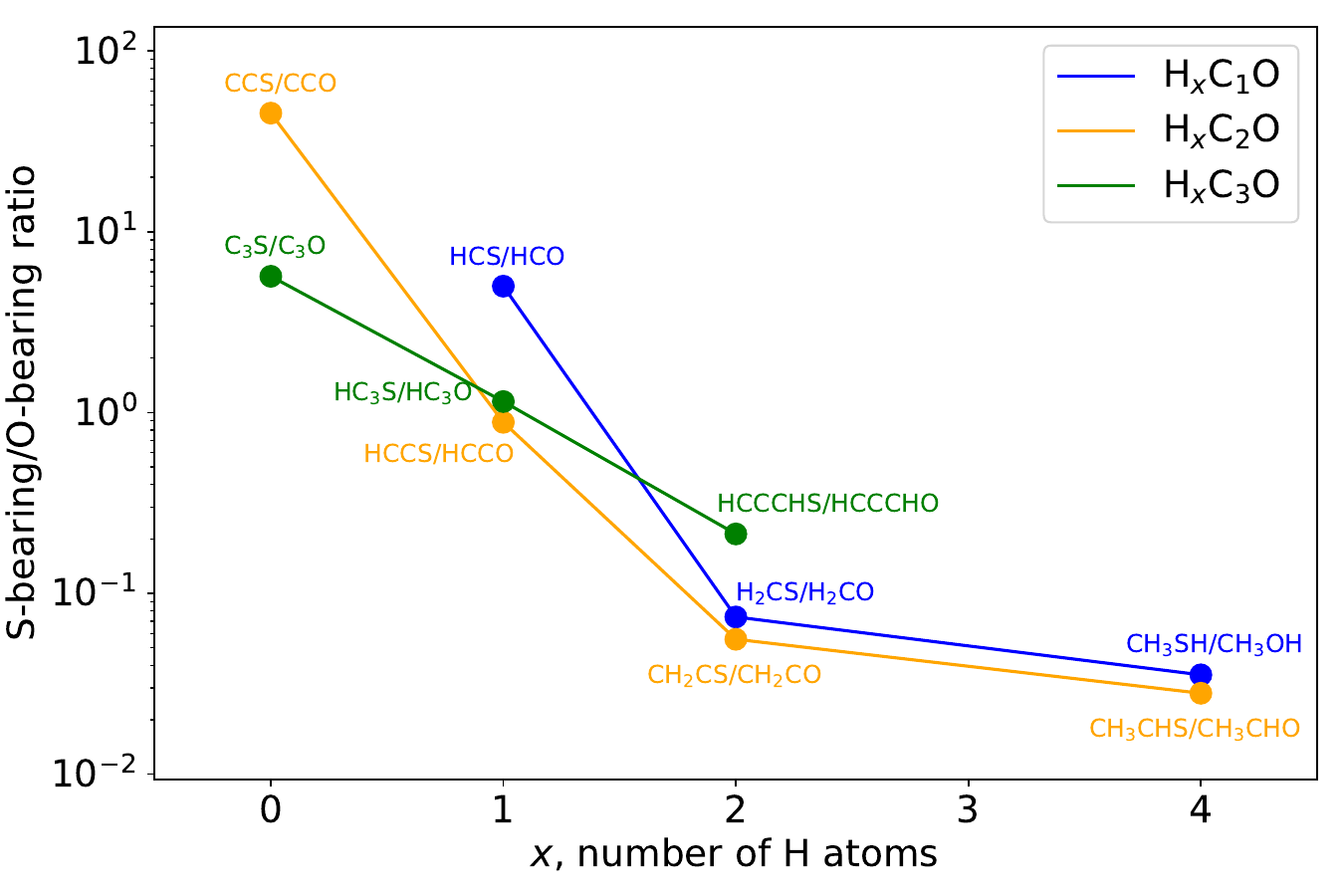}
\caption{Column density ratio between S-bearing molecule and O-bearing counterpart in \mbox{TMC-1} for the series of species H$_x$C$_1$S, H$_x$C$_2$S and H$_x$C$_3$S as a function of the number of H atoms $x$. The column density ratio CS/CO, 6.5\,$\times$\,10$^{-5}$ (see Table\,\ref{table:column_densities}), would lie outside the represented region and is not shown.} \label{fig:ratios}
\end{figure}

Many of the sulfur-bearing molecules with a detected oxygen counterpart still show remarkable differences in how their abundances behave. For example, the unsaturated carbon chains C$_2$S, C$_3$S, and C$_5$S are significantly more abundant than their oxygen analogues \citep{Cernicharo2021b,Cernicharo2021d}. Nonetheless, as these carbon chains become hydrogenated, the behavior changes. We can make a meaningful comparison between sulfur and oxygen by focusing on the series of partially hydrogenated species H$_x$C$_n$S and H$_x$C$_n$O (with $n$\,=\,1, 2, 3, and $x$\,=\,0-4), for which there are extensive observational constraints. In \mbox{TMC-1}, the molecules with $n$\,=\,1 CS, HCS, H$_2$CS, CH$_3$SH\footnote{Methyl mercaptan (CH$_3$SH) is observed in \mbox{TMC-1} through the $A$ component of the 2$_{1,1}$-1$_{1,0}$ line at 50058.809 MHz. The corresponding $E$ component at 50599.288 MHz lies outside the frequency coverage of QUIJOTE. We assumed a rotational temperature of 9 K and a source size diameter of 80\,$''$ to derive the column density value given in Table\,\ref{table:column_densities}.}, with $n$\,=\,2 CCS, HCCS, CH$_2$CS, CH$_3$CHS, and with $n$\,=\,3 C$_3$S, HC$_3$S, and HCCCHS have their oxygen counterparts detected. As seen in Table\,\ref{table:column_densities}, in some cases the oxygenated molecule is more abundant than the sulfurated one, while in other cases it is the contrary. However, a clear trend emerges when looking at the column density ratio between the S- and O-bearing species as a function of the degree of hydrogenation\footnote{A notable exception to the trend shown in Fig.\,\ref{fig:ratios} concerns the CS/CO column density ratio, which takes a value of 6.5\,$\times$\,10$^{-5}$ in \mbox{TMC-1}.}. Figure\,\ref{fig:ratios} shows that S-bearing molecules H$_x$C$_n$S with a low degree of hydrogenation ($x$\,$<$\,1) are more abundant than their corresponding oxygen counterpart, but as the degree of hydrogenation increases, the abundance of S-bearing molecules with respect to their oxygen counterparts drops. In general, for moderate to high degrees of hydrogenation ($x$\,$>$\,1), O-bearing molecules are more abundant than their corresponding sulfur analogues. A possible interpretation is that hydrogenation, whether it occurs in the gas phase or on grain surfaces, is less efficient for S-bearing molecules than for O-bearing ones, although this is probably a too simplistic view of the rather complex network of reactions involving these sulfur hydrogenated molecules \citep{Oba2018,Molpeceres2022b,Shingledecker2022}. Given the differentiation between the chemistry of sulfur and oxygen, it is not surprising that hydrogenated O-bearing complex organic molecules detected in \mbox{TMC-1}, such as C$_2$H$_3$OH, C$_2$H$_5$OH, CH$_3$OCH$_3$, HCOOCH$_3$, C$_2$H$_3$CHO, CH$_3$COCH$_3$, and C$_2$H$_5$CHO \citep{Agundez2021,Agundez2023a}, have no sulfur counterpart detected. If the trend shown in Fig.\,\ref{fig:ratios} holds at larger hydrogenation degrees, we would expect the C$_2$H$_5$SH/C$_2$H$_5$OH ratio to be lower than the CH$_3$CHS/CH$_3$CHO ratio of 0.028, meaning that the column density of C$_2$H$_5$SH would be smaller than 3.1\,$\times$\,10$^{10}$ cm$^{-2}$. It is worth noting that in warmer environments, observed C$_2$H$_5$SH/C$_2$H$_5$OH ratios are also rather small, $<$\,0.008 in Sgr\,B2 \citep{Muller2016}, 0.05 in Orion\,KL \citep{Kolesnikova2014}, and 0.07 in G+0.693$-$0.027 \citep{Rodriguez-Almeida2021}.

\begin{figure}
\centering
\includegraphics[angle=0,width=\columnwidth]{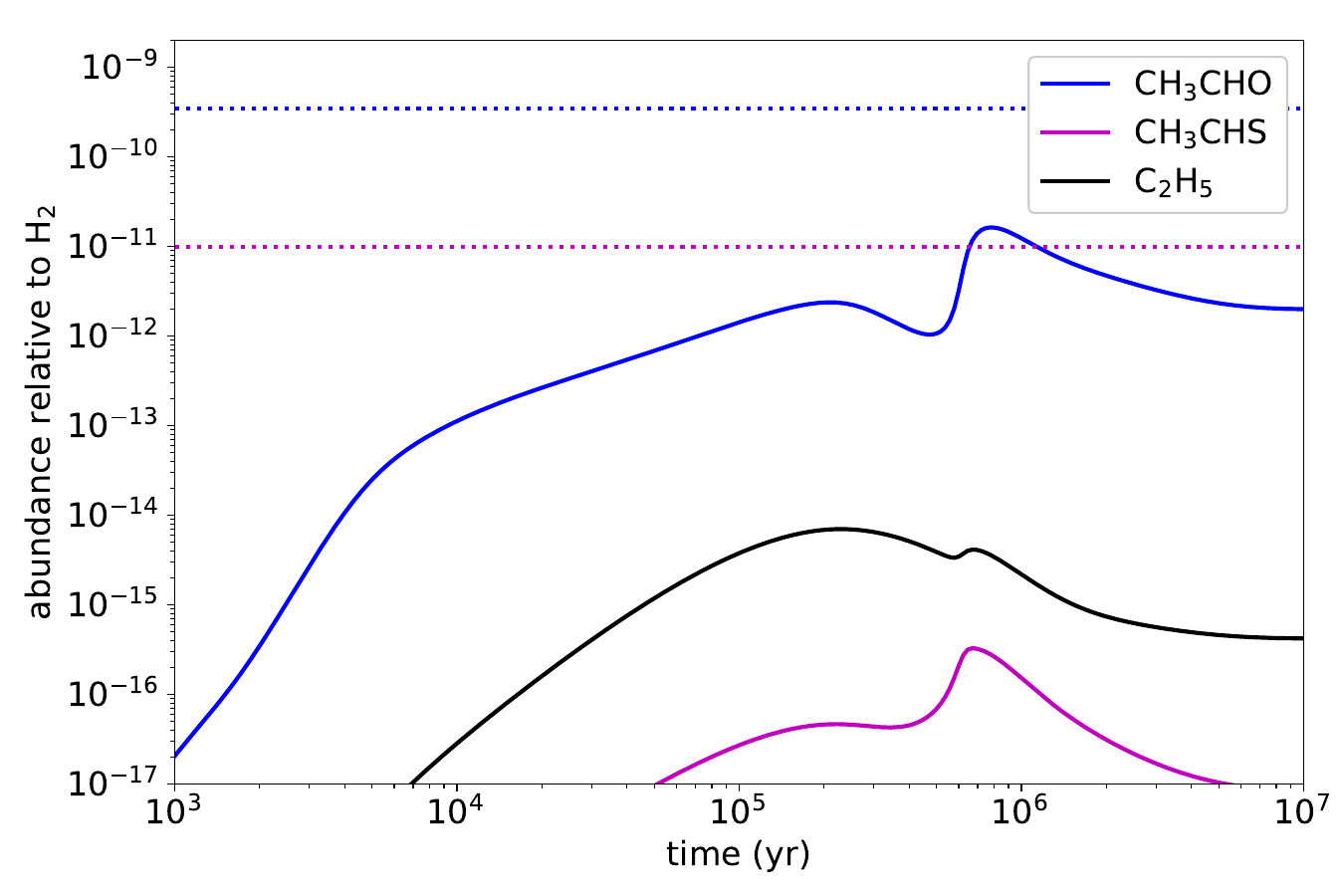}
\caption{Calculated fractional abundances of CH$_3$CHO and CH$_3$CHS, together with that of C$_2$H$_5$, are shown as a function of time. The horizontal dotted lines correspond to the values observed in \mbox{TMC-1} adopting the column densities in Table\,\ref{table:column_densities} and a column density of H$_2$ of 10$^{22}$ cm$^{-2}$ \citep{Cernicharo1987}.} \label{fig:abun}
\end{figure}

To investigate the formation of thioacetaldehyde in \mbox{TMC-1} we carried out chemical modeling calculations. We adopted a standard model of a cold dense cloud \citep{Agundez2013} and the chemical network associated to the UMIST Database for Astrochemistry 2022 \citep{Millar2024}. Thioacetaldehyde is not included in the UMIST 2022 network but its oxygen counterpart acetaldehyde is included. According to our cold dense cloud model, CH$_3$CHO is mainly formed through the dissociative recombination with electrons of protonated acetaldehyde, CH$_3$CHOH$^+$, together with the neutral-neutral reaction O + C$_2$H$_5$, which has been theoretically studied recently by \cite{Vazart2020}. In our chemical model however the calculated peak abundance of CH$_3$CHO lies around 20 times below the value observed in \mbox{TMC-1} (see Fig.\,\ref{fig:abun}). We explored whether formation pathways similar to those leading to CH$_3$CHO could explain the observed abundance of CH$_3$CHS in \mbox{TMC-1}. To that purpose, we included CH$_3$CHS as a new species in the chemical network, where we assumed that it is formed through the dissociative recombination of CH$_3$CHSH$^+$ and the neutral-neutral reaction S + C$_2$H$_5$. We adopted the same rate coefficients of the analog oxygen reactions, at the exception of S + C$_2$H$_5$, for which we carried out specific quantum chemical calculations, presented in Appendix \ref{sec:appendix_qc}. We calculated a total rate coefficient of 2.0\,$\times$\,10$^{-10}$ cm$^3$ s$^{-1}$ at 40 K, and branching ratios of 0.73 for the channel leading to H$_2$CS + CH$_3$, 0.24 for C$_2$H$_4$ + SH, 0.025 for CH$_3$CHS + H, 0.012 for CH$_2$CHSH + H, and 3.1\,$\times$\,10$^{-5}$ for cyclic CH$_2$SCH$_2$ + H. The channel leading to thioformaldehyde is therefore a minor channel, according to our calculations. The calculated abundance of CH$_3$CHS is well below the observed value (see Fig.\,\ref{fig:abun}) by more than four orders of magnitude. The reason of the strong disagreement between the chemical model and the observations is difficult to identify but it is certainly related to the poor knowledge of the chemistry of sulfur. As shown in Fig.\,\ref{fig:abun}, the calculated abundance of the ethyl radical (C$_2$H$_5$), which is a potential precursor of both CH$_3$CHO and CH$_3$CHS, is quite low and this could be at the origin of the low abundance predicted for CH$_3$CHO, although it would hardly explain the four order-of-magnitude disagreement between model and observation for CH$_3$CHS. The formation of thioacetaldehyde in \mbox{TMC-1} therefore remains unknown. We notice that the molecule has been observed as a product in solid-state experiments at 10 K in which SH, C$_2$H$_2$, and H atoms react on ices \citep{Santos2024}. The formation on ices is therefore a possibility, although it is uncertain how once formed it would desorb to the gas phase, where we observe it.

\section{Conclusions}

We reported the first detection in space of thioacetaldehyde (CH$_3$CHS), the sulfur analog of acetaldehyde (CH$_3$CHO). Thioacetaldehyde was detected toward the cyanopolyyne peak of the starless core \mbox{TMC-1} using data from the QUIJOTE program of the Yebes\,40m telescope. We derived a column density of 9.8\,$\times$\,10$^{10}$ cm$^{-2}$ for CH$_3$CHS, meaning that it is 36 times less abundant than CH$_3$CHO. A comparison of the inventory of O- and S-bearing molecules in \mbox{TMC-1} reveals a clear trend in which the abundance ratio between a sulfur-bearing molecule and its oxygen counterpart decreases with increasing hydrogenation degree, probably meaning that hydrogenation is less favored for S-bearing molecules than for O-bearing ones.

\begin{acknowledgements}

We acknowledge funding support from Spanish Ministerio de Ciencia, Innovaci\'on, y Universidades through grant PID2023-147545NB-I00 and from Consejo Superior de Investigaciones Cient\'ificas (CSIC) through project i-LINK\,23017 SENTINEL. G.M. acknowledges funding support from the Ram\'on y Cajal programme of Spanish Ministerio de Ciencia, Innovaci\'on, y Universidades through grant (RyC-2022-035442-I) and from project 20245AT016 (Proyectos Intramurales CSIC). We acknowledge the computational resources provided by the DRAGO computer cluster managed by SGAI-CSIC, and the Galician Supercomputing Center (CESGA). The supercomputer FinisTerrae III and its permanent data storage system have been funded by the Spanish Ministry of Science and Innovation, the Galician Government and the European Regional Development Fund (ERDF).

\end{acknowledgements}

\begin{appendix}

\section{Quantum chemical calculations of the gas-phase \ce{C2H5 + S} reaction} \label{sec:appendix_qc}

We tried to explore the unknown chemistry of \ce{CH3CHS} by establishing a parallelism with the chemistry of \ce{CH3CHO}. In our gas-phase model, and as mentioned in the main text, two main reactions contribute to the abundance of \ce{CH3CHO}, namely:

\begin{align}
\ce{CH3CHOH+ + e- \rightarrow CH3CHO + H} \\
\ce{C2H5 + O \rightarrow CH3CHO + H}.
\end{align}
Moving to the S-bearing counterparts, the first reaction is difficult to model, as it requires not only to know the (currently unknown) formation routes of \ce{CH3CHSH+}, but also the complete electronic structure of the cation to formulate a rate for the decay into the \ce{CH3CHS + H} channel. An easier task involves the study of the \ce{C2H5 + S \rightarrow CH3CHS + H} gas phase reaction, that we studied using quantum chemical calculations. The stationary points of the \ce{C2H5 + O} reaction were reported in \citet{Vazart2020}. We note that in their work, \citet{Vazart2020} also consider the reaction \ce{CH3OH + CH}, but we discarded this route for \ce{CH3CHS} because the chemical model indicates that even if it is rapid, its contribution to the formation of \ce{CH3CHS} would be around four orders of magnitude lower than that of the reaction \ce{C2H5 + S}.

\begin{figure*}
\centering
\includegraphics[angle=0,width=0.80\textwidth]{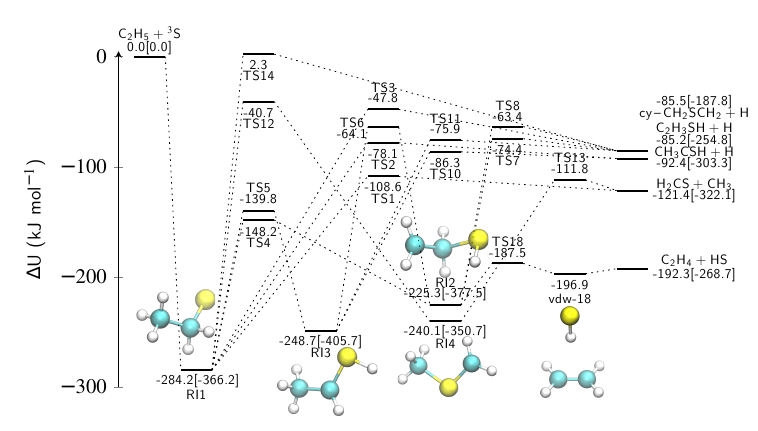}
\caption{Energy diagram for the \ce{C2H5 + S} reaction. Values in square brackets indicate energies for the same well in the \ce{C2H5 + O} reaction studied in \cite{Vazart2020}. Insets correspond to the structures at the potential wells. We note that the products cy-\ce{CH2SCH2 + H} and \ce{C2H3SH + H} are almost isoenergetic.} \label{fig:pes}
\end{figure*}

To obtain the global rate constants for the \ce{C2H5 + S} reaction we reproduced \citet{Vazart2020} calculations in the sulfur atom case. The level of theory for our calculations is selected to be exactly the same as in \citet{Vazart2020}, which facilitates any further comparison. This level of theory is CCSD(T)//aug-cc-pVTZ/B2PLYP(D3BJ)/aug-cc-pVTZ \citep{Raghavachari1989,Kendall1992, Grimme2006,Grimme2010,Grimme2011}. Zero point vibrational energies used to correct the electronic energies are also harmonic as in \citet{Vazart2020}. Finally, the choice of the code to compute the reaction energetics is made to also be consistent with \citet{Vazart2020}, i.e., Gaussian16 \citep{Frisch2016}. 

The stationary points that we obtained are gathered in Figure \ref{fig:pes}. Many points are different compared to the Potential Energy Surface (PES) of the oxygen analog reaction studied by \cite{Vazart2020}, being the most important ones the several permutations found for the most stable products of the reaction. For example, \citet{Vazart2020} report \ce{CH3} elimination, i.e., \ce{H2CO + CH3}, as the most stable reaction product, whereas in the case of sulfur, \ce{C2H4 + SH} is the preferred product. Other main changes include the permutation of the stabilities between RI1--RI3 or RI4--RI2, which in turn has an effect on the global rate constants, or the different products for TS14, for example. Moreover, one of the characteristic exit channels for the reaction in \citet{Vazart2020}, \ce{C2H3 + H2O}, was not found in the sulfur counterpart (\ce{C2H3 + H2S}). To confirm this last finding, in addition to test calculations using the equivalent structure with -O we performed additional tests performing a global minima, transition state and dissociation channel search with the Single Component Artificial Force Induced Reaction (SC-AFIR) method \citep{Maeda2013,Maeda2018} for RI2 and RI3 (the wells conducive to such channel). We used the GRRM code \citep{GRRM23} interfaced with ORCA v.5.0.3 \citep{Neese2020}. The search was performed with a model collision parameter $\gamma$ of 400 kJ mol$^{-1}$ and the computationally cheaper BHLYP(D3BJ)/6-31+G* model chemistry \citep{Becke1993,Grimme2010,Grimme2011}. Despite the intensive search, neither the dissociation channel \ce{C2H3 + H2S} nor their associated TS were found in our exploration.

\begin{table}[t]
\begin{center}
\caption{Values of the fitting parameters for the $V(r)$ attractive potential used in the barrierless association steps of the \ce{C2H5 +S} reaction. The fit is performed in the range 12-4\xspace\AA.}
\label{tab:barrierless}
\begin{tabular}{ccc}
\toprule
Reaction Step & $C_{6}$ [a.u.] & $n$ \\
\bottomrule
\ce{C2H5 + S -> C2H5S} & 469.3 & 6 \\
vdw-18\ce{-> C2H4 + HS} & 30.5 & 5 \\
\bottomrule
\end{tabular}
\end{center}
\end{table}

\begin{figure}
\centering
\includegraphics[angle=0,width=\linewidth]{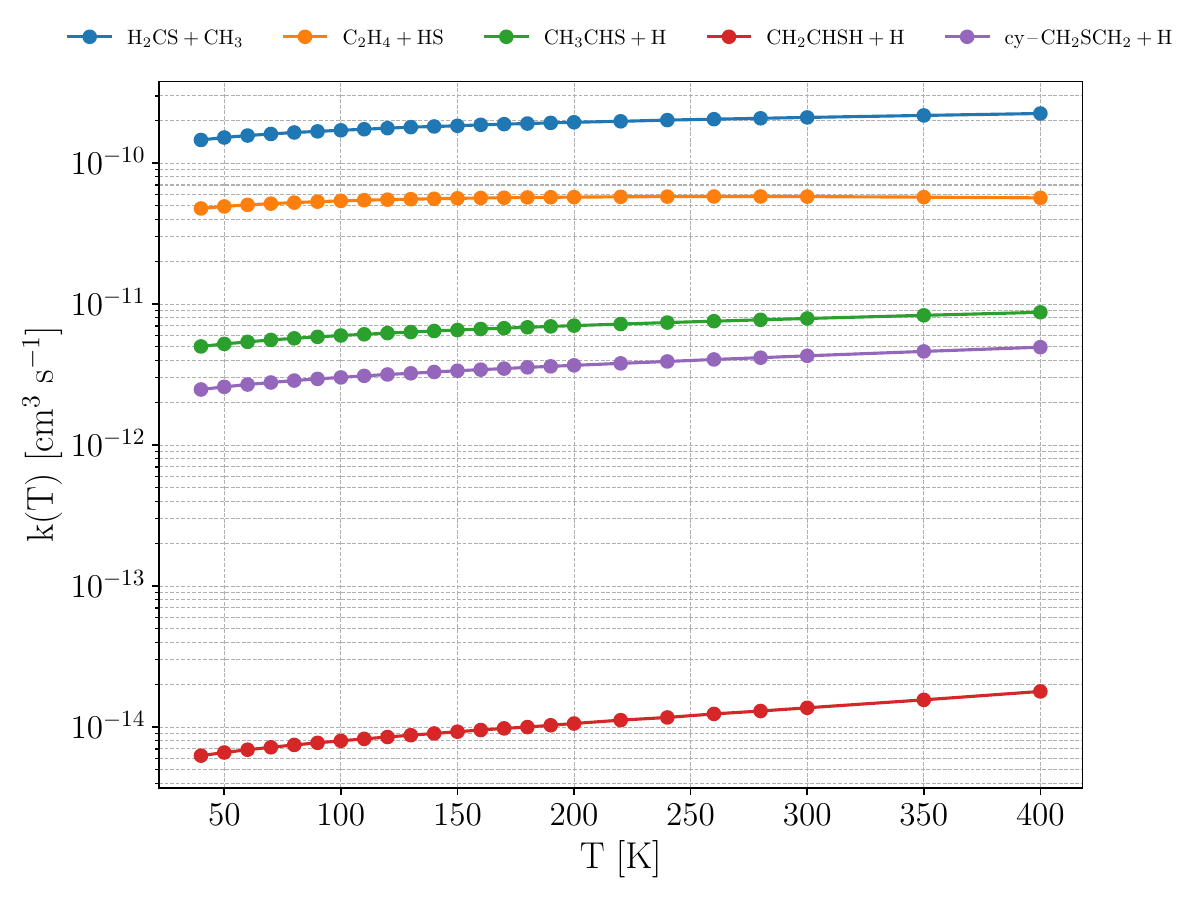}
\caption{Reaction rate constants for reaction \ce{C2H5 + S}.} \label{fig:rates}
\end{figure}            

The differences in the energetic profiles reflect on the kinetic behavior of the system. Therefore we computed global reaction rate constants for each channel using master equation computations employing RRKM theory to model unimolecular processes with a barrier and classical capture theory in the case of barrierless associations. The latter is obtained by fitting the long range interacion potential to a curve of the type $V(r)=-C_{6}/r^{n}$ functional form, with $C_{6}$ and $n$ as fitting parameters gathered in Table \ref{tab:barrierless}.  Tunneling corrections via an Eckart barrier and symmetry numbers were included, where applicable. The kinetic simulations were performed using the MESS code \citep{Georgievskii2013}. The temperature dependent rate constants are shown in Figure \ref{fig:rates}. As it can be seen from the graph, even though the \ce{C2H4 + HS} exit channel is the most stable one, it is not the dominant one owing to the competition in RI1 between isomerization to RI2 or RI4 where the latter dominates. Therefore the most dominant channel is \ce{H2CS + CH3} as in the case of the O- bearing molecule, followed by \ce{C2H4 + HS} in a variable proportion within the order of magnitude. The rest of the products, including the molecule subject of this study, \ce{CH3CHS}, are minor products of this reaction. This indicates that other pathways need to be explored to explain the presence of this molecule in \mbox{TMC-1}. Nevertheless, fitting parameters for easy inclusion of this reaction in astrochemical models, along with branching ratios at 40 K are given for every reaction channel in Table\,\ref{tab:arrhenius}.

\begin{table}[t]
\begin{center}
\caption{Modified Arrhenius parameters\,$^{a}$ for the fit of the global rate constants considered in this section over the temperature range 40-400 K. Branching ratios  (BR) are calculated from the ratio between an individual rate constant and the total loss rate at 40 K (obtained from the exact values and not from the fit). The A(B) notation indicates A$^{B}$.}
\label{tab:arrhenius}
\resizebox{\columnwidth}{!}{\begin{tabular}{ccccc}
\toprule
Reaction & $\alpha$  & $\beta$ & $\gamma$ & BR\\
\bottomrule
\ce{C2H5 + S \rightarrow H2CS + CH3} & 2.1(-10) & 2.0(-1)& -2.5(0) & 0.73 \\
\ce{C2H5 + S \rightarrow C2H4 + HS} &  5.9(-11) & -9.1(-3) & 1.1(1) & 0.24  \\
\ce{C2H5 + S \rightarrow CH3CHS + H} & 7.7(-12) & 3.2(-1) & -9.0(0) &  0.025 \\
\ce{C2H5 + S \rightarrow CH2CHSH + H} & 4.1(-12) & 4.4(-1) & -1.6(1)& 0.012 \\
\ce{C2H5 + S \rightarrow cy-CH2SCH2 + H} & 1.2(-14) & 8.0(-1) & -4.1(1) & 3.1(-5)\\
\bottomrule
\end{tabular}}
\tablenoteb{$^a$\,The modified Arrhenius expression reads:\\ $k = \alpha \left(\dfrac{T}{300\,\text{K}}\right)^\beta \exp{\left(-\dfrac{\gamma}{T}\right)}$. Units for the parameters are: $\alpha$ - cm$^{3}$ s$^{-1}$, $\beta$ - dimensionless, $\gamma$ - K.}
\end{center}
\end{table}

\end{appendix}

\end{document}